\documentstyle[preprint,aps,rotating]{revtex}
\begin{document} 
\tightenlines
\input epsf
\title{\begin{flushright}\rm\small HUB-EP-99/68
\end{flushright}
Exclusive semileptonic $\bbox{B}$ decays to radially excited 
$\bbox{D}$ mesons}
\author{D. Ebert$^1$, R. N. Faustov$^2$ and V. O. Galkin$^1$\footnote{On 
leave of absence from the Russian Academy of Sciences,
Scientific Council for Cybernetics,
Vavilov Street 40, Moscow 117333, Russia.}} 
\address{$^1$Institut f\"ur Physik, Humboldt--Universit\"at zu Berlin,
Invalidenstr.110, D-10115 Berlin, Germany\\
$^2$Russian Academy of Sciences, Scientific Council for
Cybernetics, Vavilov Street 40, Moscow 117333, Russia}
\maketitle
\begin{abstract}
Exclusive semileptonic $B$ decays to radially excited charmed mesons
are investigated at the first order of the heavy quark expansion.
The arising leading and subleading Isgur-Wise functions are calculated
in the framework of the relativistic quark model. It is found that
the $1/m_Q$ corrections play an important role and substantially
modify results. An interesting interplay between different corrections
is found. As a result the branching ratio for the $B\to D'e\nu$ decay
is essentially increased by $1/m_Q$ corrections, while the one for
$B\to D^*{'}e\nu$ is only slightly influenced by them.
\end{abstract}

\section{Introduction}
The investigation of  semileptonic decays of $B$ mesons to excited
charmed mesons represents a task interesting both from the experimental and
theoretical point of view. The current experimental data on the semileptonic
$B$ decays to the ground state $D$ mesons indicate that a substantial part  
($\approx 40\%$) of the inclusive semileptonic $B$ decays should go to 
excited $D$ meson states. First experimental data on some 
exclusive $B$ decay channels to excited charmed mesons are becoming 
available now \cite{cleo,aleph,opal} and more data are expected in near
future. Thus the comprehensive theoretical study of these decays is 
necessary. The presence of the heavy 
quark in the initial and final meson states in these decays considerably
simplifies their theoretical description. A good starting point for this
analysis is the infinitely heavy quark limit, $m_Q\to\infty$ \cite{iw}. 
In this limit the heavy quark symmetry arises, which strongly
reduces the number of independent weak form factors \cite{iw1}.
The heavy quark mass and spin then decouple and all meson properties are
determined by light-quark degrees of freedom alone. This leads to
a considerable reduction of the number of independent form factors
which are necessary for the description of heavy-to-heavy semileptonic
decays. For example, in this limit only one form factor is necessary 
for the semileptonic
$B$ decay to $S$-wave $D$ mesons (both for the ground state and its
radial excitations), while the decays to $P$ states require two form
factors \cite{iw1}. It is important to note that the 
heavy quark symmetry requires
that in the infinitely heavy quark limit matrix elements between
a $B$ meson and an excited $D$ meson should vanish at the point of
zero recoil of the final excited charmed meson in the rest frame of the $B$
meson. In the case of $B$ decays to radially excited charmed mesons
this is the result of the orthogonality of radial parts of wave functions,
while for the decays to orbital excitations this is the consequence of
orthogonality of their angular parts. However, some of the $1/m_Q$ corrections
to these decay matrix elements can give nonzero contributions at zero 
recoil. As a result the role of these  corrections could be considerably
enhanced, since the kinematical range for $B$ decays to excited states is
a rather small region around zero recoil. 
Recent calculations of semileptonic $B$ decays to
orbitally excited ($P$-wave) charmed mesons with the account of the $1/m_Q$
corrections support this observation \cite{llsw,orb}. Our calculations
\cite{orb} in the framework of the relativistic quark model show
that  some  rates of $B$ decays to orbitally excited charmed mesons receive
contributions from  first order $1/m_Q$ corrections approximately of
the same value as a leading order contribution. In this paper we extend
our analysis to $B$ decays to radially excited $D$ mesons.

Our relativistic quark model is  based on the quasipotential approach in
quantum field
theory with a specific choice  of the quark-antiquark interaction
potential. It provides
a consistent scheme for the  calculation of all relativistic corrections
at a given $v^2/c^2$ 
order and allows for the heavy  quark $1/m_Q$ expansion. In preceding
papers 
we applied this model to the calculation of the mass spectra of 
orbitally and radially excited states of heavy-light mesons \cite{egf},
as well as to a description of weak decays of $B$ mesons to ground state
heavy and light mesons \cite{fgm,efg}. The heavy quark expansion for the ground
state heavy-to-heavy semileptonic transitions \cite{fg} has been found to be 
in agreement with model-independent predictions of the HQET.

The paper is organized as follows. In Sec.~II we cary out the heavy quark 
expansion for the weak decay matrix elements between a $B$ meson and radially 
excited charmed meson states up to the first order in $1/m_Q$  
using  heavy quark 
effective theory (HQET). In our analysis we follow HQET derivations for 
the matrix elements between ground states \cite{luke,n} and for 
the matrix element between a $B$ meson and orbitally excited charmed meson
\cite{llsw}, as well as a general analysis of these matrix elements in
Ref.~\cite{mr}. In Sec.~III we describe our relativistic quark model.
The heavy quark expansion for decay matrix elements is then carried out up to
the first order $1/m_Q$ corrections
and compared to the model-independent HQET results in Secs.~IV-V. We
determine the leading and subleading Isgur-Wise functions and give
our predictions for decay branching ratios in the heavy quark limit and with
the account of $1/m_Q$ corrections. The electron spectra for the
considered decays are also presented. Section ~IV contains our conclusions.    

   \section{Decay matrix elements and the heavy quark
      expansion}
      
The matrix elements of the vector current ($J^V_\mu=\bar c\gamma_\mu b$) and
axial vector current ($J^A_\mu=\bar c\gamma_\mu\gamma_5 b$) between
$B$ and radially excited $D'$ or $D^{*}{'}$ mesons can be parameterized by six
hadronic form factors:
 \begin{eqnarray}\label{ff}
 {\langle D'(v')| \bar c\gamma^\mu b |B(v)\rangle  
\over\sqrt{m_{D'}m_B}}
  &=& h_+ (v+v')^\mu + h_- (v-v')^\mu , \cr\cr
  \langle D'(v')| \bar c\gamma^\mu b \gamma_5 |B(v)\rangle 
  &=& 0, \cr\cr
  {\langle D^*{'}(v',\epsilon)| \bar c\gamma^\mu b |B(v)\rangle  
\over\sqrt{m_{D^*{'}}m_B}}
  &=& i h_V \varepsilon^{\mu\alpha\beta\gamma} 
  \epsilon^*_\alpha v'_\beta v_\gamma ,\cr\cr
  {\langle D^*{'}(v',\epsilon)| \bar c\gamma^\mu\gamma_5 b |B(v)\rangle  
\over\sqrt{m_{D^*{'}}m_B}}
  &=& h_{A_1}(w+1) \epsilon^{* \mu} 
   -(h_{A_2} v^\mu + h_{A_3} v'^\mu) (\epsilon^*\cdot v) ,
   \end{eqnarray}
where $v~(v')$ is the four-velocity of the $B~(D^{(*)}{'})$ meson,
$\epsilon^\mu$ is a  polarization vector  of the final vector
charmed meson, and the form factors $h_i$  are dimensionless 
functions of the product of velocities $w=v\cdot v'$. The double  
differential decay rates expressed in terms of
the form factors are
\begin{eqnarray}\label{ddr}
{{\rm d}^2\Gamma_{D'}\over {\rm d}w{\rm d}\!\cos\theta} &=& 
  3\Gamma_0 r^{3} (w^2-1)^{3/2} \sin^2\theta 
  \Big[ (1+r)h_+ - (1-r) h_- \Big]^2 ,\cr \cr
{{\rm d}^2\Gamma_{D^{*}{'}}\over {\rm d}w{\rm d}\!\cos\theta} &=& 
  3\Gamma_0 r^{*3} \sqrt{w^2-1} \bigg\{ \sin^2\theta
  \Big[ (w-r^*) h_{A_1}+(w^2-1) (h_{A_3}+r^* h_{A_2}) \Big]^2 \cr \cr
&& + (1-2r^*w+r^{*2}) \Big[(1+\cos^2\theta) [h_{A_1}^2 + (w^2-1) h_V^2] 
  - 4\cos\theta \sqrt{w^2-1} h_{A_1} h_V \Big] \bigg\} ,
\end{eqnarray}
where $\Gamma_0 = {G_F^2\,|V_{cb}|^2\,m_B^5 /(192\pi^3)}$, 
$r=m_{D'}/m_B$ , $r^*=m_{D^{*}{'}}/m_B$ and $\theta$ is  the 
angle between the charged lepton and the 
charmed meson in the rest frame of the virtual $W$ boson.

Now we expand the form factors $h_i$ in powers of $1/m_Q$ up to first
order and relate the coefficients in this expansion to universal Isgur-Wise
functions. This is achieved by evaluating the matrix elements of the
effective current operators arising from the HQET expansion of the weak
currents. For simplicity we limit our analysis to the leading order in
$\alpha_s$ and use the trace formalism \cite{falk}. 
Following Ref.~\cite{llsw}, we introduce the matrix
\begin{equation}
H_v = \frac{1+\not\! v}{2} \Big[ P_v^{*\mu} \gamma_\mu 
  - P_v \gamma_5 \Big],  \label{Hdef}
\end{equation} 
composed from the fields $P_v$ and $P_v^{*\mu}$ that destroy mesons in
the $j^{P}=\frac12^-$ doublet \footnote{Here $j$ is the total light quark 
angular momentum, and the superscript $P$ denotes the meson parity.} 
with four-velocity $v$. At leading
order of the heavy quark expansion ($m_Q\to\infty$) the matrix elements of
the weak current between the  ground and radially excited states 
destroyed  by the fields in $H_v$ and $H'_v$, respectively, 
are given by
\begin{equation}\label{lo}
\bar c \Gamma b \to \bar h^{(c)}_{v'}\Gamma h^{(b)}_v = \xi^{(n)}(w)
  {\rm Tr} \Big\{\bar H'_{v'} \Gamma H_v \Big\},
\end{equation} 
where $h_v^{(Q)}$ is the heavy quark field in the effective theory.
The leading order Isgur-Wise function $\xi^{(n)}(w)$ vanishes at the 
zero recoil ($w=1$) of the final meson for any $\Gamma$, because of the
heavy quark symmetry and the orthogonality of the radially excited state
wave function with respect to the ground state one.
 
At first order of the $1/m_Q$ expansion there are contributions from
the corrections to the HQET Lagrangian 
 \begin{eqnarray}\label{lcor}
\delta{\cal L}& = & \frac1{2m_Q}{\cal L}_{1,v}^{(Q)}
\equiv\frac1{2 m_Q} \Big[ O_{{\rm kin},v}^{(Q)} +
  O_{{\rm mag},v}^{(Q)} \Big], \\
O_{{\rm kin},v}^{(Q)}& =& \bar h_v^{(Q)} (iD)^2 h_v^{(Q)}, \qquad
O_{{\rm mag},v}^{(Q)} = \bar h_v^{(Q)}
  \frac{g_s}2 \sigma_{\alpha\beta} G^{\alpha\beta} h_v^{(Q)}\nonumber
\end{eqnarray}  
and from the tree-level matching of the weak current operator onto
effective theory which contains a
covariant derivative $D^{\lambda}=\partial^{\lambda}-ig_st_aA_a^{\lambda}$
 \begin{equation}\label{ccor}
\bar c \Gamma b \to \bar h_{v'}^{(c)} 
  \bigg( \Gamma - \frac i{2m_c} \overleftarrow {\not\!\! D} \Gamma  
  + \frac i{2m_b} \Gamma \overrightarrow {\not\!\! D} 
  \bigg) h_v^{(b)}. 
\end{equation}
 The matrix elements of the latter operators can be parameterized as
\begin{eqnarray}\label{cur}
\bar h^{(c)}_{v'} i\overleftarrow D_{\lambda} \Gamma h^{(b)}_v &=&
  {\rm Tr}\Big\{ \xi_{\lambda}^{(c)} 
  \bar H_{v'}\Gamma H_v \Big\} , \cr
\bar h^{(c)}_{v'} \Gamma i\overrightarrow D_{\lambda} h^{(b)}_v &=&
  {\rm Tr} \Big\{  \xi^{(b)}_{\lambda} 
  \bar H_{v'} \Gamma H_v \Big\}.  
\end{eqnarray}
The most general form for $\xi^{(Q)}_{\lambda}$ is \cite{luke}
\begin{equation}
\xi^{(Q)}_{\lambda}=\xi^{(Q)}_{+}(v+v')_{\lambda}+\xi^{(Q)}_{-}
(v-v')_{\lambda}-\xi^{(Q)}_3\gamma_{\lambda}.
\end{equation}
The equation of motion for the heavy quark, $i(v\cdot D)h^{(Q)}=0$, yields
the relations between the form factors $\xi^{(Q)}_i$
\begin{eqnarray}\label{rel1}
\xi_{+}^{(c)}(1+w)+\xi_{-}^{(c)}(w-1)+\xi_3^{(c)}&=&0\cr
\xi_{+}^{(b)}(1+w)-\xi_{-}^{(b)}(w-1)+\xi_3^{(b)}&=&0.
\end{eqnarray}
The additional relations can be obtained from the  momentum conservation
and the definition of the heavy quark fields $h_v^{(Q)}$, which lead to
the equation 
$ i\partial_\nu(\bar h_{v'}^{(c)}\Gamma\,h_v^{(b)}) = 
  (\bar\Lambda v_\nu-\bar\Lambda^{(n)}v'_\nu)\bar h_{v'}^{(c)}\Gamma 
h_v^{(b)}$, implying that
\begin{equation}\label{cnst}
{\xi}^{(c)}_{\lambda} + {\xi}^{(b)}_{\lambda}
  = (\bar\Lambda v_\lambda-\bar\Lambda^{(n)}v'_\lambda) \xi^{(n)}.
\end{equation}
Here $\bar\Lambda(\bar\Lambda^{(n)})=M(M^{(n)})-m_Q$ is the difference between
the heavy ground state (radially excited) meson and heavy quark masses in 
the limit $m_Q\to\infty$. This equation results in the following relations
\begin{eqnarray}\label{rel2}
\xi^{(c)}_{+}+\xi^{(b)}_{+}+\xi^{(c)}_{-}+\xi^{(b)}_{-}
&=&\bar\Lambda\xi^{(n)},\cr
\xi^{(c)}_{+}+\xi^{(b)}_{+}-\xi^{(c)}_{-}-\xi^{(b)}_{-}
&=&-\bar\Lambda^{(n)}\xi^{(n)},\cr
\xi_3^{(c)}+\xi_3^{(b)}&=&0.
\end{eqnarray}
The relations (\ref{rel1}) and (\ref{rel2}) can be used to 
express the functions
$\xi_{-,+}^{(Q)}$ in terms of $\tilde\xi_3(\equiv\xi_3^{(b)}=-\xi_3^{(c)})$ and
the leading order function $\xi^{(n)}$:
\begin{eqnarray}\label{corc}
\xi^{(c)}_{-}&=&\left(\frac{\bar\Lambda^{(n)}}2+
\frac12\frac{\bar\Lambda^{(n)}-\bar\Lambda}{w-1}\right)\xi^{(n)},\cr\cr
\xi^{(b)}_{-}&=&\left(\frac{\bar\Lambda}2-
\frac12\frac{\bar\Lambda^{(n)}-\bar\Lambda}{w-1}\right)\xi^{(n)},\cr\cr
\xi^{(c)}_{+}&=&\left(-\frac{\bar\Lambda^{(n)}}2+
\frac12\frac{\bar\Lambda^{(n)}+\bar\Lambda}{w+1}\right)\xi^{(n)}+
\frac1{w+1}\tilde\xi_3,\cr\cr
\xi^{(b)}_{+}&=&\left(\frac{\bar\Lambda}2-
\frac12\frac{\bar\Lambda^{(n)}+\bar\Lambda}{w+1}\right)\xi^{(n)}-
\frac1{w+1}\tilde\xi_3.
\end{eqnarray}

The matrix elements of the $1/m_Q$ corrections resulting from insertions
of higher-dimension operators of the HQET Lagrangian (\ref{lcor}) 
have the structure \cite{luke}
\begin{eqnarray}\label{corl}
i \int {\rm d}x\, T\left\{ {\cal L}_{1,v'}^{(c)}(x) 
  \left[ \bar h_{v'}^{(c)} \Gamma h_{v}^{(b)} \right](0) \right\} 
  &=& 2\chi^{(c)}_1 {\rm Tr} \left\{\bar H_{v'} \Gamma H_v \right\}+
2{\rm Tr}\left\{ \chi_{\alpha\beta}^{(c)}
  \bar H_{v'} i\sigma^{\alpha\beta} \frac{1+\not \! v'}2 
  \Gamma H_v \right\},\cr\cr
i \int {\rm d}x\, T\left\{ {\cal L}_{1,v}^{(b)}(x) 
  \left[ \bar h_{v'}^{(c)} \Gamma h_{v}^{(b)} \right](0) \right\} 
  &=& 2\chi^{(b)}_1 {\rm Tr} \left\{\bar H_{v'} \Gamma H_v \right\}+
2{\rm Tr}\left\{ \chi_{\alpha\beta}^{(b)}
  \bar H_{v'}\Gamma \frac{1+\not\! v}2 i\sigma^{\alpha\beta}  
   H_v \right\}.
\end{eqnarray}
The corrections coming from the kinetic energy term $O_{\rm kin}$ 
do not violate spin symmetry and, hence, the corresponding functions
$\chi_1^{(Q)}$ effectively correct the leading order function $\xi^{(n)}$.
The chromomagnetic operator $O_{\rm mag}$, on the other hand, explicitly
violates spin symmetry. The most general decomposition of the tensor
form factor $\chi_{\alpha\beta}^{(Q)}$ is \cite{luke,n}
\begin{eqnarray}
\chi_{\alpha\beta}^{(c)}&=&\chi_2^{(c)}v_{\alpha}\gamma_{\beta} 
-\chi_3^{(c)}i\sigma_{\alpha\beta},\cr
\chi_{\alpha\beta}^{(b)}&=&\chi_2^{(b)}v'_{\alpha}\gamma_{\beta} 
-\chi_3^{(b)}i\sigma_{\alpha\beta}.
\end{eqnarray}

The functions $\chi_i^{(b)}$ contribute to the decay form factors (\ref{ff})
only in the linear combination $\chi_b=2\chi_1^{(b)}-4(w-1)\chi_2^{(b)}
+12\chi_3^{(b)}$. Thus five independent functions $\tilde\xi_3$, $\chi_b$ and
$\tilde\chi_i(\equiv \chi_i^{(c)})$, as well as two mass parameters
$\Lambda$ and $\Lambda^{(n)}$ are necessary to describe first order $1/m_Q$
corrections to  matrix elements of $B$ meson decays to radially excited
$D$ meson states. The resulting structure of the decay form factors is
\begin{eqnarray}\label{cff}
h_{+}&=&\xi^{(n)}+\varepsilon_c\left[2\tilde\chi_1-4(w-1)\tilde\chi_2+
12\tilde\chi_3\right]+\varepsilon_b\chi_b,\cr\cr
h_{-}&=&\varepsilon_c\left[2\tilde\xi_3-\left(\bar\Lambda^{(n)}+
\frac{\bar\Lambda^{(n)}-\bar\Lambda}{w-1}\right)\xi^{(n)}\right]
- \varepsilon_b\left[2\tilde\xi_3-\left(\bar\Lambda-
\frac{\bar\Lambda^{(n)}-\bar\Lambda}{w-1}\right)\xi^{(n)}\right],\cr\cr
h_V&=&\xi^{(n)}+\varepsilon_c\left[2\tilde\chi_1-4\tilde\chi_3+
\left(\bar\Lambda^{(n)}+\frac{\bar\Lambda^{(n)}
-\bar\Lambda}{w-1}\right)\xi^{(n)}\right]\cr\cr
&&+\varepsilon_b\left[\chi_b+
\left(\bar\Lambda-\frac{\bar\Lambda^{(n)}-\bar\Lambda}{w-1}\right)\xi^{(n)}
-2\tilde\xi_3\right],\cr\cr
h_{A_1}&=&\xi^{(n)}+\varepsilon_c\left[2\tilde\chi_1-4\tilde\chi_3+
\frac{w-1}{w+1}\left(\bar\Lambda^{(n)}+\frac{\bar\Lambda^{(n)}
-\bar\Lambda}{w-1}\right)\xi^{(n)}\right]\cr\cr
&&+\varepsilon_b\left\{\chi_b+
\frac{w-1}{w+1}\left[
\left(\bar\Lambda-\frac{\bar\Lambda^{(n)}-\bar\Lambda}{w-1}\right)\xi^{(n)}
-2\tilde\xi_3\right]\right\},\cr\cr
h_{A_2}&=&\varepsilon_c\left\{4\tilde\chi_2-\frac2{w+1}\left[
\left(\bar\Lambda^{(n)}+\frac{\bar\Lambda^{(n)}
-\bar\Lambda}{w-1}\right)\xi^{(n)}+\tilde\xi_3\right]\right\},\cr\cr
h_{A_3}&=&\xi^{(n)}+\varepsilon_c\left[2\tilde\chi_1-4\tilde\chi_2-
4\tilde\chi_3+\frac{w-1}{w+1}\left(\bar\Lambda^{(n)}+\frac{\bar\Lambda^{(n)}
-\bar\Lambda}{w-1}\right)\xi^{(n)}-\frac2{w+1}\tilde\xi_3\right]\cr\cr
&&+\varepsilon_b\left[\chi_b+
\left(\bar\Lambda-\frac{\bar\Lambda^{(n)}-\bar\Lambda}{w-1}\right)\xi^{(n)}
-2\tilde\xi_3\right],
\end{eqnarray}
where $\varepsilon_Q=1/(2m_Q)$.

In the following sections we apply the relativistic quark model to  the
calculation
of leading and subleading Isgur-Wise functions.

\section{Relativistic quark model}

In the quasipotential approach, a meson is described by the wave
function of the bound quark-antiquark state, which satisfies the
quasipotential equation \cite{3} of the Schr\"odinger type~\cite{4}:
\begin{equation}
\label{quas}
{\left(\frac{b^2(M)}{2\mu_{R}}-\frac{{\bf
p}^2}{2\mu_{R}}\right)\Psi_{M}({\bf p})} =\int\frac{d^3 q}{(2\pi)^3}
 V({\bf p,q};M)\Psi_{M}({\bf q}),
\end{equation}
where the relativistic reduced mass is
\begin{equation}
\mu_{R}=\frac{M^4-(m^2_q-m^2_Q)^2}{4M^3}.
\end{equation}
Here $m_{q,Q}$ are the masses of light
and heavy quarks, and ${\bf p}$ is their relative momentum.  
In the center-of-mass system the relative momentum squared on mass shell 
reads
\begin{equation}
{b^2(M) }
=\frac{[M^2-(m_q+m_Q)^2][M^2-(m_q-m_Q)^2]}{4M^2}.
\end{equation}

The kernel 
$V({\bf p,q};M)$ in Eq.~(\ref{quas}) is the quasipotential operator of
the quark-antiquark interaction. It is constructed with the help of the
off-mass-shell scattering amplitude, projected onto the positive
energy states. An important role in this construction is played 
by the Lorentz-structure of the confining quark-antiquark interaction
in the meson.  In 
constructing the quasipotential of the quark-antiquark interaction 
we have assumed that the effective
interaction is the sum of the usual one-gluon exchange term and the mixture
of vector and scalar linear confining potentials.
The quasipotential is then defined by
\cite{mass}  
\begin{eqnarray}
\label{qpot}
V({\bf p,q};M)&=&\bar{u}_q(p)\bar{u}_Q(-p){\cal V}({\bf p}, {\bf
q};M)u_q(q)u_Q(-q)\cr \cr
&=&\bar{u}_q(p)
\bar{u}_Q(-p)\Bigg\{\frac{4}{3}\alpha_sD_{ \mu\nu}({\bf
k})\gamma_q^{\mu}\gamma_Q^{\nu}\cr\cr
& & +V^V_{\rm conf}({\bf k})\Gamma_q^{\mu}
\Gamma_{Q;\mu}+V^S_{\rm conf}({\bf
k})\Bigg\}u_q(q)u_Q(-q),
\end{eqnarray}
where $\alpha_s$ is the QCD coupling constant, $D_{\mu\nu}$ is the
gluon propagator in the Coulomb gauge
and ${\bf k=p-q}$; $\gamma_{\mu}$ and $u(p)$ are 
the Dirac matrices and spinors
\begin{equation}
\label{spinor}
u^\lambda({p})=\sqrt{\frac{\epsilon(p)+m}{2\epsilon(p)}}
{1\choose \frac{\bbox{\sigma p}}{\epsilon(p)+m}}\chi^\lambda
\end{equation}
with $\epsilon(p)=\sqrt{{\bf p}^2+m^2}$.
The effective long-range vector vertex is
given by
\begin{equation}
\Gamma_{\mu}({\bf k})=\gamma_{\mu}+
\frac{i\kappa}{2m}\sigma_{\mu\nu}k^{\nu},
\end{equation}
where $\kappa$ is the Pauli interaction constant characterizing the
nonperturbative anomalous chromomagnetic moment of quarks. Vector and
scalar confining potentials in the nonrelativistic limit reduce to
\begin{equation}
V^V_{\rm conf}(r)=(1-\varepsilon)(Ar+B),\qquad
V^S_{\rm conf}(r) =\varepsilon (Ar+B),
\end{equation}
reproducing 
\begin{equation}
V_{\rm conf}(r)=V^S_{\rm conf}(r)+
V^V_{\rm conf}(r)=Ar+B,
\end{equation}
where $\varepsilon$ is the mixing coefficient. 

The quasipotential for the heavy quarkonia,
expanded in $v^2/c^2$, can be found in Refs.~\cite{mass,pot} and for
heavy-light mesons in \cite{egf}.
All the parameters of
our model, such as quark masses, parameters of the linear confining potential,
mixing coefficient $\varepsilon$ and anomalous
chromomagnetic quark moment $\kappa$, were fixed from the analysis of
heavy quarkonia masses \cite{mass} and radiative decays \cite{gf}. 
The quark masses
$m_b=4.88$ GeV, $m_c=1.55$ GeV, $m_s=0.50$ GeV, $m_{u,d}=0.33$ GeV and
the parameters of the linear potential $A=0.18$ GeV$^2$ and $B=-0.30$ GeV
have the usual quark model values.  
In Ref.~\cite{fg} we have considered the expansion of  the matrix
elements of weak heavy quark currents between pseudoscalar and vector
meson ground states up to the second order in inverse powers of the heavy quark
masses. It has been found that the general structure of the leading,
first,
and second order $1/m_Q$ corrections in our relativistic model is in accord 
with the predictions of HQET. The heavy quark symmetry and QCD impose rigid 
constraints on the parameters of the long-range potential in our model.
The analysis
of the first order corrections \cite{fg} allowed us to fix the value of the 
Pauli interaction
constant $\kappa=-1$. The same value of $\kappa$  was found previously
from  the fine splitting of heavy quarkonia ${}^3P_J$- states \cite{mass}. 
\footnote{It has been known for a long time 
that the correct reproduction of the
spin-dependent part of the quark-antiquark interaction requires 
either assuming  the scalar confinement or equivalently  introducing the
Pauli interaction with $\kappa=-1$ \cite{schn,mass,pot} in the vector
confinement.}
Note that the long-range chromomagnetic spin-dependent
interaction in our model
is proportional to $(1+\kappa)$ and thus vanishes for  
$\kappa=-1$ in agreement with the flux tube model \cite{buch}. The
value of the mixing 
parameter of vector and scalar confining potentials $\varepsilon=-1$
has been found from the analysis of the second order corrections \cite{fg}.
This value is very close to the one determined from radiative decays of
heavy quarkonia \cite{gf}.   

In order to calculate the exclusive semileptonic decay rate of the $B$ meson, 
it is necessary to determine the corresponding matrix element of the 
weak current between meson states.
In the quasipotential approach,  the matrix element of the weak current
$J^W=\bar c\gamma_\mu(1-\gamma^5)b$ 
between a $B$ meson and a radially excited $D^{(*)}{'}$ meson takes
 the form \cite{f}
\begin{equation}\label{mxet} 
\langle D^{(*)}{'} \vert J^W_\mu (0) \vert B\rangle
=\int \frac{d^3p\, d^3q}{(2\pi )^6} \bar \Psi_{D^{(*)}{'}}({\bf
p})\Gamma _\mu ({\bf p},{\bf q})\Psi_B({\bf q}),\end{equation}
where $\Gamma _\mu ({\bf p},{\bf
q})$ is the two-particle vertex function and  $\Psi_{B,D^{(*)}{'}}$ are the
meson wave functions projected onto the positive energy states of
quarks and boosted to the moving reference frame.
 The contributions to $\Gamma$ come from Figs.~1 and 2.\footnote{  
The contribution $\Gamma^{(2)}$ is the consequence
of the projection onto the positive-energy states. Note that the form of the
relativistic corrections resulting from the vertex function
$\Gamma^{(2)}$ is explicitly dependent on the Lorentz structure of the
$q\bar q$-interaction.} In the heavy quark limit
$m_{b,c}\to \infty$ only $\Gamma^{(1)}$ contributes, while $\Gamma^{(2)}$ 
contributes at $1/m_{Q}$ order. 
They look like
\begin{equation} \label{gamma1}
\Gamma_\mu^{(1)}({\bf
p},{\bf q})=\bar u_{c}(p_c)\gamma_\mu(1-\gamma^5)u_b(q_b)
(2\pi)^3\delta({\bf p}_q-{\bf
q}_q),\end{equation}
and
\begin{eqnarray}\label{gamma2} 
\Gamma_\mu^{(2)}({\bf
p},{\bf q})&=&\bar u_{c}(p_c)\bar u_q(p_q) \Bigl\{\gamma_{Q\mu}(1-\gamma_Q^5)
\frac{\Lambda_b^{(-)}(
k)}{\epsilon_b(k)+\epsilon_b(p_c)}\gamma_Q^0
{\cal V}({\bf p}_q-{\bf
q}_q)\nonumber \\ 
& &+{\cal V}({\bf p}_q-{\bf
q}_q)\frac{\Lambda_{c}^{(-)}(k')}{ \epsilon_{c}(k')+
\epsilon_{c}(q_b)}\gamma_Q^0 \gamma_{Q\mu}(1-\gamma_Q^5)\Bigr\}u_b(q_b)
u_q(q_q),\end{eqnarray}
where the superscripts ``(1)" and ``(2)" correspond to Figs.~1 and
2, $Q= c$ or $b$, ${\bf k}={\bf p}_c-{\bf\Delta};\
{\bf k}'={\bf q}_b+{\bf\Delta};\ {\bf\Delta}={\bf
p}_{D^{(*)}{'}}-{\bf p}_B$;
$$\Lambda^{(-)}(p)=\frac{\epsilon(p)-\bigl( m\gamma
^0+\gamma^0({\bbox{ \gamma p}})\bigr)}{ 2\epsilon (p)}.$$
Here \cite{f} 
\begin{eqnarray*} 
p_{c,q}&=&\epsilon_{c,q}(p)\frac{p_{D^{(*)}{'}}}{M_{D^{(*)}{'}}}
\pm\sum_{i=1}^3 n^{(i)}(p_{D^{(*)}{'}})p^i,\\
q_{b,q}&=&\epsilon_{b,q}(q)\frac{p_B}{M_B} \pm \sum_{i=1}^3 n^{(i)}
(p_B)q^i,\end{eqnarray*}
and $n^{(i)}$ are three four-vectors given by
$$ n^{(i)\mu}(p)=\left\{ \frac{p^i}{M},\ \delta_{ij}+
\frac{p^ip^j}{M(E+M)}\right\}, \quad E=\sqrt{{\bf p}^2+M^2}.$$

It is important to note that the wave functions entering the weak current
matrix element (\ref{mxet}) are not in the rest frame in general. For example, 
in the $B$ meson rest frame, the $D^{(*)}{'}$ meson is moving with the recoil
momentum ${\bf \Delta}$. The wave function
of the moving $D^{(*)}{'}$ meson $\Psi_{D^{(*)}{'}\,{\bf\Delta}}$ is connected 
with the $D^{(*)}{'}$ wave function in the rest frame 
$\Psi_{D^{(*)}{'}\,{\bf 0}}$ by the transformation \cite{f}
\begin{equation}
\label{wig}
\Psi_{D^{(*)}{'}\,{\bf\Delta}}({\bf
p})=D_c^{1/2}(R_{L_{\bf\Delta}}^W)D_q^{1/2}(R_{L_{
\bf\Delta}}^W)\Psi_{D^{(*)}{'}\,{\bf 0}}({\bf p}),
\end{equation}
where $R^W$ is the Wigner rotation, $L_{\bf\Delta}$ is the Lorentz boost
from the meson rest frame to a moving one, and   
the rotation matrix $D^{1/2}(R)$ in spinor representation is given by
\begin{equation}\label{d12}
{1 \ \ \,0\choose 0 \ \ \,1}D^{1/2}_{c,q}(R^W_{L_{\bf\Delta}})=
S^{-1}({\bf p}_{c,q})S({\bf\Delta})S({\bf p}),
\end{equation}
where
$$
S({\bf p})=\sqrt{\frac{\epsilon(p)+m}{2m}}\left(1+\frac{\bbox{ \alpha p}}
{\epsilon(p)+m}\right)
$$
is the usual Lorentz transformation matrix of the four-spinor.

\section{Leading and subleading Isgur-Wise functions}

Now we can perform the heavy quark expansion for the matrix elements
of $B$ decays to radially excited $D$ mesons in the framework of our model and
determine leading and subleading Isgur-Wise functions. We substitute
the vertex functions $\Gamma^{(1)}$  and $\Gamma^{(2)}$ 
given by Eqs.~(\ref{gamma1}) and (\ref{gamma2})
in the decay matrix element (\ref{mxet})  and take into account the wave
function transformation (\ref{wig}).
The resulting structure of this matrix element is
rather complicated, because it is necessary  to integrate both over  $d^3 p$
and $d^3 q$. The $\delta$ function in expression  (\ref{gamma1}) permits
us to perform
one of these integrations and thus this contribution  can be easily
calculated. The calculation  
of the vertex function $\Gamma^{(2)}$ contribution is  more difficult.
Here, instead
of a $\delta$ function, we have a complicated structure, containing the 
$Q\bar q$ interaction potential in the meson. 
However, we can expand this contribution in inverse
powers of heavy ($b,c$) quark masses and then use the  quasipotential
equation in
order to perform one of the integrations in the current matrix element.  
We carry out 
the heavy quark expansion up to first order in $1/m_Q$. It is easy to see  
that the vertex
function $\Gamma^{(2)}$ contributes already at  the subleading order of  
the $1/m_Q$
expansion. Then we compare the arising  decay matrix elements with
the form factor decomposition (\ref{ff}) and determine the corresponding
form factors. We find that, for the chosen values of our model parameters  
(the mixing coefficient of vector and scalar confining potential 
$\varepsilon=-1$  and the Pauli
constant $\kappa=-1$), the resulting structure  at leading
and subleading order in $1/m_Q$ coincides with the model-independent  
predictions of HQET  given by Eq.~(\ref{cff}). We get the following
expressions for leading and subleading Isgur-Wise functions:
\begin{eqnarray}
\label{xi}  
\xi^{(1)}(w)&=&\left(\frac{2}{w+1}\right)^{1/2}\int\frac{d^3 p}
{(2\pi)^3}\bar\psi^{(0)}_{D^{(*)}{'}}\!\!\left({\bf p}+
\frac{2\epsilon_q}{M_{D^{(*)}{'}}(w+1)}
{\bf \Delta}\right)
\psi^{(0)}_B({\bf p}),\\ \cr
\label{xi3}
\tilde\xi_3(w)&=&\left(\frac{\bar\Lambda^{(1)}+\bar\Lambda}2-m_q+
\frac16\frac{\bar\Lambda^{(1)}-\bar\Lambda}{w-1}\right)\left(1+
\frac23\frac{w-1}{w+1}\right)\xi^{(n)}(w),\\ \cr
\label{chi1}
\tilde\chi_1(w)&\cong &\frac1{20}\frac{w-1}{w+1}\frac{\bar\Lambda^{(1)}
-\bar\Lambda}{w-1}\xi^{(1)}(w)\cr \cr
&&+\frac{\bar\Lambda^{(1)}}2
\left(\frac{2}{w+1}\right)^{1/2}\int\frac{d^3 p}
{(2\pi)^3}\bar\psi^{(1)si}_{D^{(*)}{'}}\!\!\left({\bf p}+
\frac{2\epsilon_q}{M_{D^{(*)}{'}}(w+1)}
{\bf \Delta}\right)
\psi^{(0)}_B({\bf p}),\\ \cr
\label{chi2}
\tilde\chi_2(w)&\cong& -\frac1{12}\frac{1}{w+1}\frac{\bar\Lambda^{(1)}
-\bar\Lambda}{w-1}\xi^{(1)}(w),\\ \cr
\label{chi3}
\tilde\chi_3(w)&\cong& -\frac3{80}\frac{w-1}{w+1}\frac{\bar\Lambda^{(1)}
-\bar\Lambda}{w-1}\xi^{(1)}(w)\cr \cr
&&+\frac{\bar\Lambda^{(1)}}4
\left(\frac{2}{w+1}\right)^{1/2}\int\frac{d^3 p}
{(2\pi)^3}\bar\psi^{(1)sd}_{D^{(*)}{'}}\!\!\left({\bf p}+
\frac{2\epsilon_q}{M_{D^{(*)}{'}}(w+1)}
{\bf \Delta}\right)
\psi^{(0)}_B({\bf p}),\\ \cr
\label{chib}
\chi_b(w)&\cong& \bar\Lambda\left(\frac{2}{w+1}\right)^{1/2}\int\frac{d^3 p}
{(2\pi)^3}\bar\psi^{(0)}_{D^{(*)}{'}}\!\!\left({\bf p}+
\frac{2\epsilon_q}{M_{D^{(*)}{'}}(w+1)}
{\bf \Delta}\right)
\left[\psi^{(1)si}_B({\bf p})-3\psi^{(1)sd}_B({\bf p})\right].
\end{eqnarray}
Here we used the expansion for the $S$-wave meson wave function
$$\psi_M=\psi_M^{(0)}+\bar\Lambda_M\varepsilon_Q\left(\psi_M^{(1)si}
+d_M\psi_M^{(1)sd}\right)+O(1/m_Q^2),$$
where $\psi_M^{(0)}$ is the wave function in the limit $m_Q\to\infty$,
$\psi_M^{(1)si}$ and $\psi_M^{(1)sd}$ are the spin-independent and 
spin-dependent first order $1/m_Q$ corrections, $d_P=-3$ for pseudoscalar and
$d_V=1$ for vector mesons.
The symbol $\cong$ in the expressions (\ref{chi1})--(\ref{chib}) for the
subleading functions $\tilde\chi_i(w)$ means that the corrections
suppressed by an additional power of the ratio $(w-1)/(w+1)$, which is equal 
to zero  at $w=1$ and less than $1/6$ at $w_{\rm max}$, were neglected. 
Since the main contribution to the decay rate comes from the values of 
form factors  close to $w=1$, these corrections turn out to be unimportant. 
 
It is clear from the expression (\ref{xi}) that the leading order contribution
vanishes at the point of zero recoil (${\bf \Delta}=0, w=1$) of the 
final $D^{(*)}{'}$ meson, 
since the radial parts of the wave functions $\Psi_{D^{(*)}{'}}$ and $\Psi_B$
are orthogonal in the infinitely heavy quark limit. The $1/m_Q$ 
corrections to the current (\ref{corc}) also do not contribute at this 
kinematical point for the same reason.
The only nonzero contributions at $w=1$  come from corrections to the
Lagrangian~\footnote{There are no normalization conditions for 
these corrections
contrary to the decay to the ground state $D^{(*)}$ mesons, where the 
conservation of vector current requires their vanishing at
zero recoil \cite{luke}.} $\tilde\chi_1(w)$, $\tilde\chi_3(w)$ and 
$\chi_b(w)$. From Eqs.~(\ref{cff}) one can find for the form factors
contributing to the decay matrix elements at zero recoil 
\begin{eqnarray}\label{zeroc}
h_{+}(1)&=&\varepsilon_c\left[2\tilde\chi_1(1)+
12\tilde\chi_3(1)\right]+\varepsilon_b\chi_b(1),\cr
h_{A_1}(1)&=&\varepsilon_c\left[2\tilde\chi_1(1)-4\tilde\chi_3(1)\right]
+\varepsilon_b\chi_b(1).
\end{eqnarray}
Such nonvanishing contributions at zero recoil   
result from the first order $1 /m_Q$ corrections to the wave functions (see 
Eq.~(\ref{chib}) and the last terms in
Eqs.~(\ref{chi1}), (\ref{chi3})). Since the kinematically
allowed range for these decays is not broad ( $1\le w\le 
w_{\rm max}\approx 1.27$) the relative contribution to the decay rate of such 
small $1/m_Q$ corrections is substantially increased.  
Note that the terms 
$\varepsilon_Q(\bar\Lambda^{(n)}-\bar\Lambda)\xi^{(n)}(w)/(w-1)$ have
the same behaviour near $w=1$ as the leading order contribution, in contrast
to decays to the ground state $D^{(*)}$ mesons, where $1/m_Q$ corrections 
are suppressed with respect to the leading order contribution 
by the factor $(w-1)$ near this point (this result is known as Luke's 
theorem \cite{luke}). Since inclusion of first order heavy quark corrections to
$B$ decays to the ground state $D^{(*)}$ mesons results in approximately a 
10-20\% increase of decay rates \cite{fg,n}, one could expect that the 
influence of these corrections on decay rates to radially excited 
$D^{(*)}{'}$ mesons will be more essential. Our numerical analysis supports
these observations.

\section{Numerical results and predictions} 
In Table~\ref{massp} we present the masses of radially 
excited $D'$ and $D^{*}{'}$
as well as mass parameters $\bar\Lambda$ calculated in the framework of our
model \cite{egf}. Our prediction for the $D^{*}{'}$ mass is in good 
agreement with the DELPHI measurement \cite{delphi}.
Other radially excited states have not been observed yet. Thus we use 
our predictions for numerical calculations. The values of leading and
subleading Isgur-Wise functions (\ref{xi})--(\ref{chib}) and their slopes
at the point of zero recoil of the final $D^{(*)}{'}$ meson are given in
Table~\ref{vf}. 
In Fig.~3 we plot our
results for the leading order Isgur-Wise function $\xi^{(1)}(w)$ and
the current correction function $\tilde\xi_3(w)$. 
The functions $\tilde\chi_1(w)$,
$\tilde\chi_2(w)$, $\tilde\chi_3(w)$, and $\chi_b(w)$ are plotted in Fig.~4.
We see that the functions, parameterizing chromomagnetic corrections
to the HQET Lagrangian, are rather small in accord with the HQET based 
expectations. 

We can now calculate the decay  branching ratios by integrating double
differential  decay rates in Eq.~(\ref{ddr}). Our results for decay rates
both in the infinitely heavy quark limit and taking account of the 
first order $1/m_Q$ corrections as well as their ratio
$$R=\frac{{\rm Br}(B\to D^{(*)}{'}e\nu)_{{\rm with}\, 1/m_Q}}{{\rm Br}(B\to
D^{(*)}{'}e\nu)_{m_Q\to\infty}}$$
are presented in Table~\ref{tbr}. 
We find that both $1/m_Q$ corrections to decay
rates arising from corrections to HQET Lagrangian (\ref{chi1})--(\ref{chib}),
which do not vanish at zero recoil, and corrections to the current (\ref{xi3}),
(\ref{corc}), vanishing at zero recoil, give significant contributions. In
the case of $B\to D'e\nu$ decay both types of these corrections tend to
increase the decay rate leading to approximately a 75\% increase of the
$B\to D'e\nu$ decay rate. On the other hand, these corrections give opposite
contributions to the $B \to D^*{'}e\nu$ decay rate: the corrections to the
current give a negative contribution, while  corrections to the Lagrangian
give a positive one of approximately the same value. This interplay of
$1/m_Q$ corrections only slightly ($\approx 10\%$) increases the decay rate
with respect to the infinitely heavy quark limit. As a result the branching
ratio for $B\to D'e\nu$ decay exceeds the one for $B\to D^*{'}e\nu$
after inclusion of first order $1/m_Q$ corrections. In the infinitely heavy
quark mass limit we have for the ratio $Br(B\to D'e\nu)/Br(B\to D^*{'}e\nu)
=0.75$, while the account of $1/m_Q$ corrections results in the considerable 
increase of this ratio  to 1.22. 

In Table~\ref{tbr} we also present the sum of the branching ratios over first
radially excited states. Inclusion of $1/m_Q$ corrections results in
approximately a 40\% increase of this sum. We see that our model predicts
that $ 0.40\%$ of $B$ meson decays go to the first radially excited $D$
meson states. If we add this value to our prediction for $B$ decays to
the first orbitally excited states $ 1.45\%$ \cite{orb}, we 
get the value of 1.85\%. This result means that approximately 2\% of
$B$ decays should go to higher excitations.

In Figs.~5 and 6 we plot the electron spectra $(1/\Gamma_0) ({\rm d}\Gamma/{\rm
d}y)$ for $B\to D'e\nu$ and
$B\to D^{*}{'}e\nu$ decays. Here $y=2E_e/m_B$ is the  rescaled lepton
energy. These
differential decay rates can be easily obtained from  double differential
decay rates (\ref{ddr}),  using the relation
$y=1-rw-r\sqrt{w^2-1}\cos\theta$ and then integrating in $w$ over  
$[(1-y)^2+r^2]/[2r(1-y)]<w<(1+r^2)/(2r)$. We present
our  results both in the heavy quark limit $m_Q\to\infty$ (dashed curves)
and with the inclusion of first order $1/m_Q$ corrections (solid curves).
From Fig.~6 we see that inclusion of $1/m_Q$ corrections significantly
changes the shape of electron spectrum for $B\to D^*{'}e\nu$ decay. 
The maximum is considerably shifted to higher lepton energies.

\section{Conclusions}
In this paper we have carried out the heavy quark expansion for the decay
matrix elements of weak currents between the $B$ meson and radially excited
$D^{(*)}{'}$ meson states up to first order. It is found that five 
additional functions of the product of velocities  $w$ are necessary 
to parameterize 
first order $1/m_Q$ corrections. One of these functions $\tilde\xi_3(w)$
arises from corrections to the weak current. The other four functions
$\chi_i(w)$ ($i=1\dots 3,b$) parameterize corrections to the HQET Lagrangian.
The leading order function  vanishes at the point of zero recoil ($w=1$)
of the final $D^{(*)}{'}$ meson due to the heavy quark symmetry and 
orthogonality of the radial parts of meson wave functions in the heavy
quark limit. The contributions to the decay matrix elements coming from
the corrections to the current also vanish at zero recoil for the same
reason. Thus the only nonzero contributions to the weak decay matrix elements
at $w=1$  come from the corrections to the Lagrangian $\tilde\chi_{1,3,b}$
(see Eq.~(\ref{zeroc})), since there is no condition requiring them to
vanish at this point as in the case of $B$ decays to ground state $D$ mesons.

Then we apply the relativistic quark model for the consideration of
semileptonic  $B\to D^{(*)}{'}e\nu$ decays. It is found 
that our model correctly
reproduces the structure of decay matrix elements found from the heavy quark
symmetry analysis. This allows us to determine the leading
and subleading Isgur-Wise functions for this transitions. We find that both the
relativistic transformation of the meson wave function from the rest frame
to the moving one as well as the first order $1/m_Q$ corrections to meson
wave functions give essential contributions to the subleading order functions.
Thus, the account for corrections to the wave functions gives contributions 
to decay matrix elements which do not vanish at zero recoil. These 
contributions turn out to be rather small numerically. However, their role is
considerably increased since the kinematical range for $B$ decays
to radially excited $D^{(*)}{'}$ mesons is  small and thus the leading
order contribution, vanishing at zero recoil, is suppressed. Another important
contribution to decay rates comes from the terms $\varepsilon_Q(\Lambda^{(1)}
-\Lambda)\xi^{(1)}/(w-1)$ originating from the corrections to the current.
These terms turn out to be numerically important. We find an 
interesting interplay  of these two types of $1/m_Q$ corrections. They
contribute to the  $B\to D'e\nu$ decay rate with the same sign, but
their contributions to the $B\to D^*{'}e\nu$ rate have opposite signs. As a
result the former decay rate is substantially (1.75 times) increased by 
the inclusion of first order $1/m_Q$ corrections while there is only
a slight (1.1 times) increase of the latter decay rate. This leads to
the  increase in the ratio $Br(B\to D'e\nu)/Br(B\to D^*{'}e\nu)$
from 0.75 in the heavy quark limit to 1.22, when  first order
$1/m_Q$ corrections are taken into account. Finally, we find that
the semileptonic $B$ decays to first radial excitations of $D$ mesons 
acquire in total 0.4\% of the $B$ decay rate.      
    
\acknowledgements
We thank M. Beneke,  J.G. K\"orner and V.I. Savrin 
for useful discussions.
V.O.G. gratefully acknowledges the warm hospitality
of the colleagues in the particle theory group of the Humboldt-University
extended to him during his stay there.
R.N.F and V.O.G. were supported in part by the {\it Deutsche
Forschungsgemeinschaft} under contract Eb 139/1-3.

\begin{table}
\caption{Masses of radially excited $D^{(*)}{'}$ mesons and the mass parameters
$\bar\Lambda$ in our model.}
\label{massp}
\begin{tabular}{ccccc}
Parameter&State&Value (GeV) \cite{egf}&Exp. (GeV) \cite{delphi}\\
\hline
$m_{D'}$& $D'(2S_0)$& 2.579 &   \\
$m_{D^{*}{'}}$& $D^*{'}(2S_1)$&2.629 & 2.637(9) \\
$\bar\Lambda$& $B,D(1S)$&0.51& \\
$\bar\Lambda^{(1)}$& $D(2S)$& 0.94& \\
$m_{D_s'}$& $D_s'(2S_0)$& 2.670 &  \\
$m_{D_s^{*}{'}}$&$D_s^{*}{'}(2S_1)$& 2.716& \\
$\bar\Lambda_s$& $B_s,D_s(1S)$&0.61& \\
$\bar\Lambda_s^{(1)}$& $D_s(2S)$& 1.05& \\
\end{tabular}
\end{table}

\begin{table}
\caption{ Leading and subleding Isgur-Wise functions and their slopes
$\left.\rho^2_{\xi_i}=-\frac1{\xi_i}\frac{\partial}{\partial w}
\xi_i\right|_{w=1}$ and $\left.\rho^2_{\chi_i}=
-\frac1{\chi_i}\frac{\partial}{\partial w}\chi_i\right|_{w=1}$ at zero recoil.
We factored out $(w-1)$ from the leading order form factor and defined 
$\xi^{(1)}(w)=(w-1)\Xi(w)$. The values of the functions $\tilde\xi_3(1)$, 
$\chi_i(1)$ are given in units $(\bar\Lambda^{(1)}+\bar\Lambda)/2$.}
\label{vf}
\begin{tabular}{ccccccc}
 & $\Xi(w)$ &$\tilde\xi_3(w)$ &$\tilde\chi_1(w)$ &$\tilde\chi_2(w)$
&$\tilde\chi_3(w)$&$\chi_b(w)$\\
\hline
Value at $w=1$& 2.2 &0.21 &0.18 &$-0.054$& $-0.023$& $-0.098$\\
Slope at $w=1$& 2.6 &$-3.3$&1.9&3.1& 1.0& 2.1\\
\end{tabular}
\end{table}

\begin{table}
\caption{Decay rates $\Gamma$ (in units  of $|V_{cb}/0.04|^2\times
10^{-15}$ GeV) 
and branching ratios BR (in \%) for  $B$ ($B_s$) decays to radially
excited $D^{(*)}{'}$ ($D_s^{(*)}{'}$) mesons in the
infinitely heavy quark
mass limit and taking account of first  order $1/m_Q$ corrections. 
$\Sigma (B\to D^{(*)}{'}e\nu)$ and $\Sigma (B_s\to D_s^{(*)}{'}e\nu)$
represent the sum over the channels. 
$R$ is a ratio
of branching ratios taking account of  $1/m_Q$ corrections to branching
ratios in
the infinitely heavy quark mass limit.  }
\label{tbr}
\begin{tabular}{cccccc}
   &\multicolumn{2}{c}{$m_Q\to\infty$}&\multicolumn{2}{c}{With $1/m_Q$}\\
Decay& $\Gamma$ & Br& $\Gamma$ & Br &$R$ \\
\hline
$B\to D'e\nu$&0.53&0.12  & 0.92 & 0.22& 1.74\\
$B\to D^{*}{'}e\nu$&0.70&0.17 & 0.78 & 0.18& 1.11\\
$\Sigma (B\to D^{(*)}{'}e\nu)$ & 1.23& 0.29&1.70&0.40&1.37\\
\hline
$B_s\to D'_{s}e\nu$&0.66&0.16 & 1.18 & 0.28&1.80 \\
$B_s\to D_{s}^*{'}e\nu$&0.86&0.20 & 0.95 & 0.22&1.10 \\
$\Sigma (B_s\to D_s^{(*)}{'}e\nu)$& 1.52&0.36&2.13&0.50&1.40\\
\end{tabular}
\end{table}

\begin{figure}
\unitlength=0.9mm
\large
\begin{picture}(150,150)
\put(10,100){\line(1,0){50}}
\put(10,120){\line(1,0){50}}
\put(35,120){\circle*{8}}
\multiput(32.5,130)(0,-10){2}{\begin{picture}(5,10)
\put(2.5,10){\oval(5,5)[r]}
\put(2.5,5){\oval(5,5)[l]}\end{picture}}
\put(5,120){$b$}
\put(5,100){$\bar q$}
\put(5,110){$B$}
\put(65,120){$c$}
\put(65,100){$\bar q$}
\put(65,110){$D^{(*)}{'}$}
\put(43,140){$W$}
\put(0,85){\small FIG. 1. Lowest order vertex function $\Gamma^{(1)}$
contributing to the current matrix element (25). }
\put(10,20){\line(1,0){50}}
\put(10,40){\line(1,0){50}}
\put(25,40){\circle*{8}}
\put(25,40){\thicklines \line(1,0){20}}
\multiput(25,40.5)(0,-0.1){10}{\thicklines \line(1,0){20}}
\put(25,39.5){\thicklines \line(1,0){20}}
\put(45,40){\circle*{2}}
\put(45,20){\circle*{2}}
\multiput(45,40)(0,-4){5}{\line(0,-1){2}}
\multiput(22.5,50)(0,-10){2}{\begin{picture}(5,10)
\put(2.5,10){\oval(5,5)[r]}
\put(2.5,5){\oval(5,5)[l]}\end{picture}}
\put(5,40){$b$}
\put(5,20){$\bar q$}
\put(5,30){$B$}
\put(65,40){$c$}
\put(65,20){$\bar q$}
\put(65,30){$D^{(*)}{'}$}
\put(33,60){$W$}
\put(90,20){\line(1,0){50}}
\put(90,40){\line(1,0){50}}
\put(125,40){\circle*{8}}
\put(105,40){\thicklines \line(1,0){20}}
\multiput(105,40.5)(0,-0.1){10}{\thicklines \line(1,0){20}}
\put(105,39,5){\thicklines \line(1,0){20}}
\put(105,40){\circle*{2}}
\put(105,20){\circle*{2}}
\multiput(105,40)(0,-4){5}{\line(0,-1){2}}
\multiput(122.5,50)(0,-10){2}{\begin{picture}(5,10)
\put(2.5,10){\oval(5,5)[r]}
\put(2.5,5){\oval(5,5)[l]}\end{picture}}
\put(85,40){$b$}
\put(85,20){$\bar q$}
\put(85,30){$B$}
\put(145,40){$c$}
\put(145,20){$\bar q$}
\put(145,30){$D^{(*)}{'}$}
\put(133,60){$W$}
\put(0,5){\small FIG. 2. Vertex function $\Gamma^{(2)}$
taking the quark interaction into account. Dashed lines correspond  }
\put(0,0) {\small to the effective potential ${\cal V}$ in 
(\ref{qpot}). Bold lines denote the negative-energy part of the quark}
\put(0,-5){\small propagator. }

\end{picture}

\end{figure}
\setcounter{figure}2

\begin{figure}
\centerline{
\begin{turn}{-90}
\epsfxsize=14 cm
\epsfbox{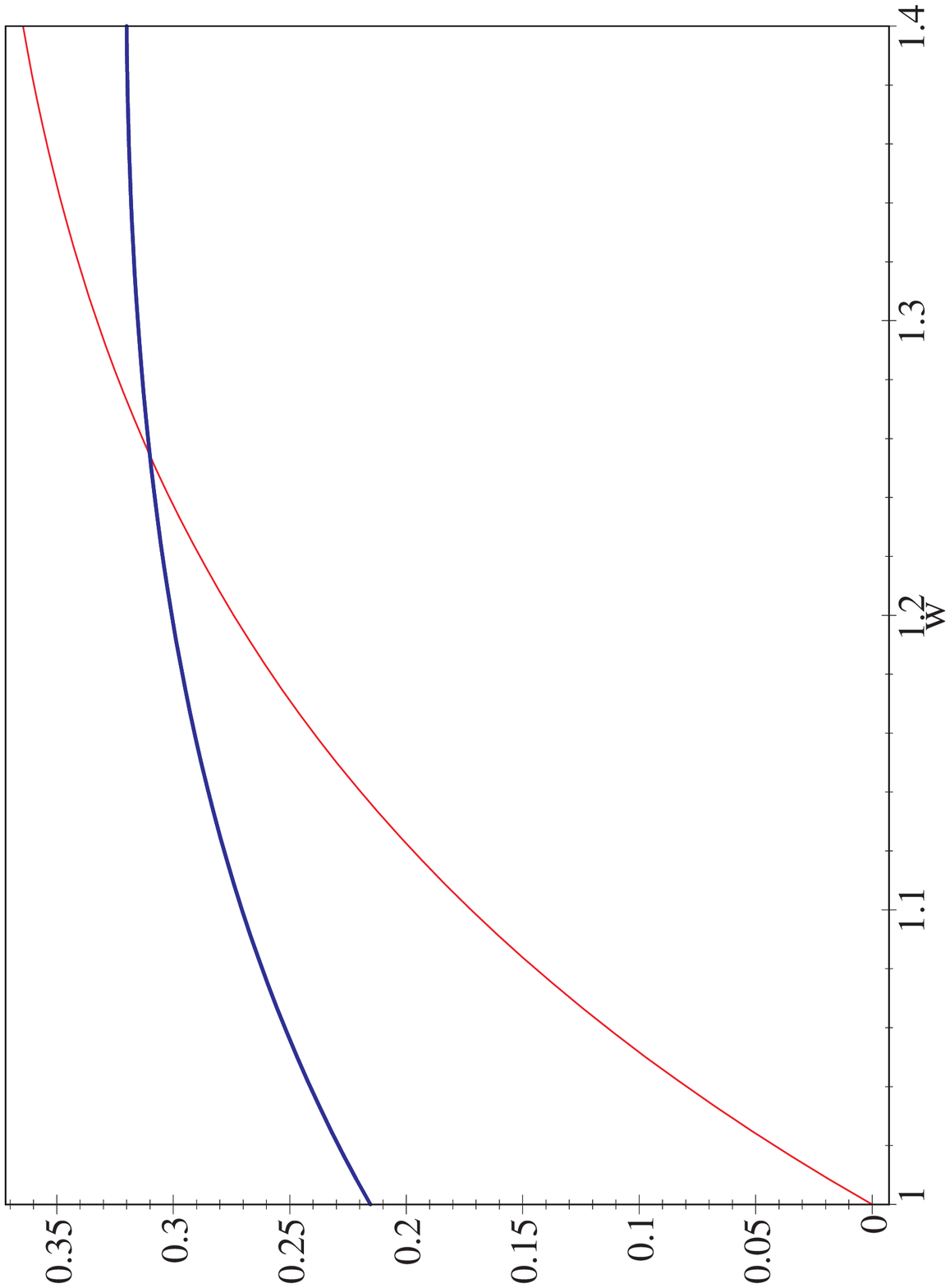}
\end{turn}
}

\caption{Isgur-Wise functions $\xi^{(1)}(w)$ (solid curve) 
and $\tilde\xi_3(w)$
(bold curve, in units $(\bar\Lambda^{(1)}+\bar\Lambda)/2$) 
for the $B\to D^{(*)}{'}e\nu$
decay.}
\end{figure}

\begin{figure}
\centerline{
\begin{turn}{-90}
\epsfxsize=14 cm
\epsfbox{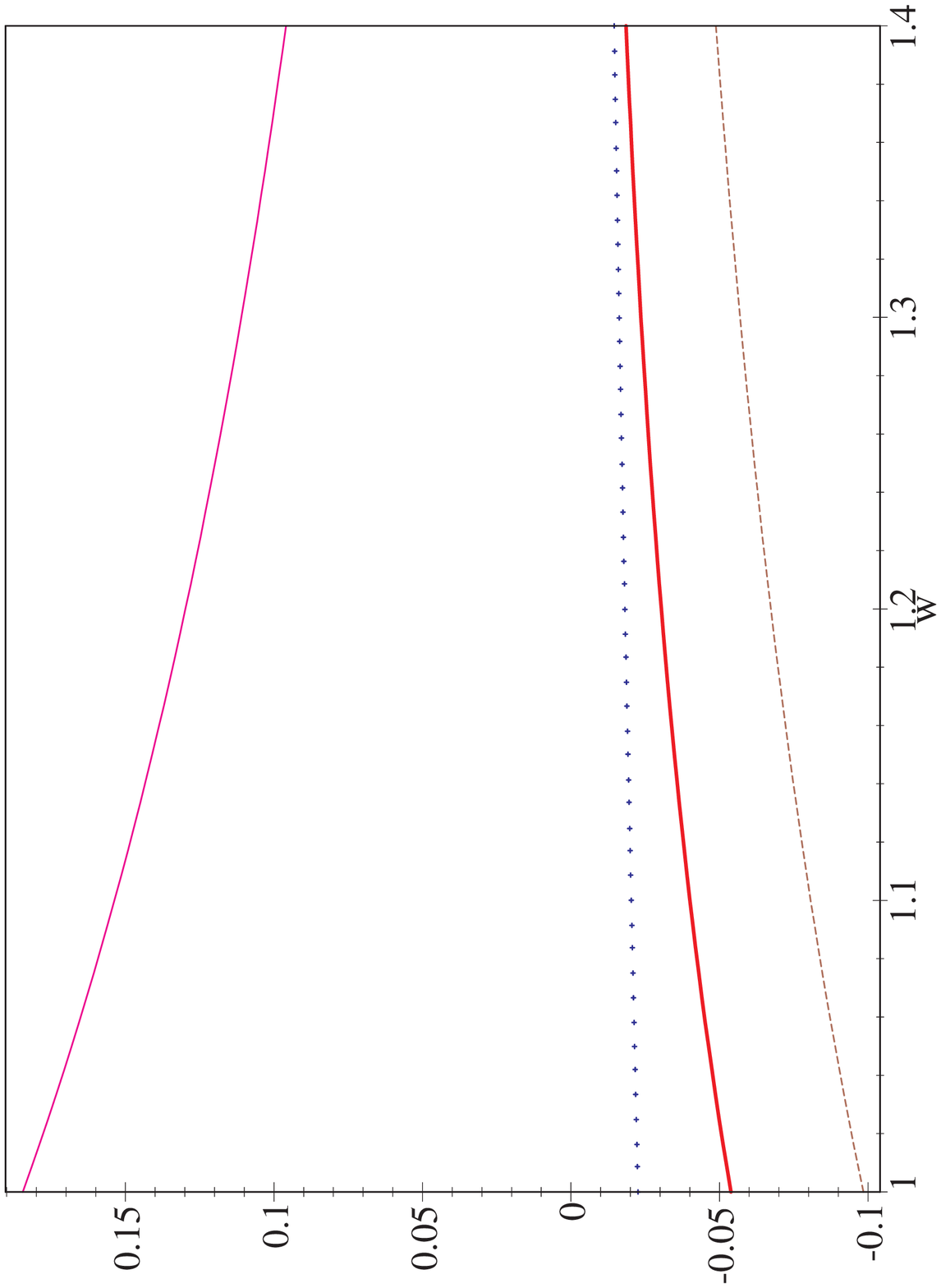}
\end{turn}
}

\caption{Isgur-Wise functions $\tilde\chi_1(w)$ (solid curve), 
$\tilde\chi_2(w)$ (bold curve),
$\tilde\chi_3(w)$ (dotted curve), and $\chi_b(w)$ (dashed curve) 
for the $B\to D^{(*)}{'}e\nu$
decay in units $(\bar\Lambda^{(1)}+\bar\Lambda)/2$.}

\end{figure}

\begin{figure}
\centerline{
\begin{turn}{-90}
\epsfxsize=14 cm
\epsfbox{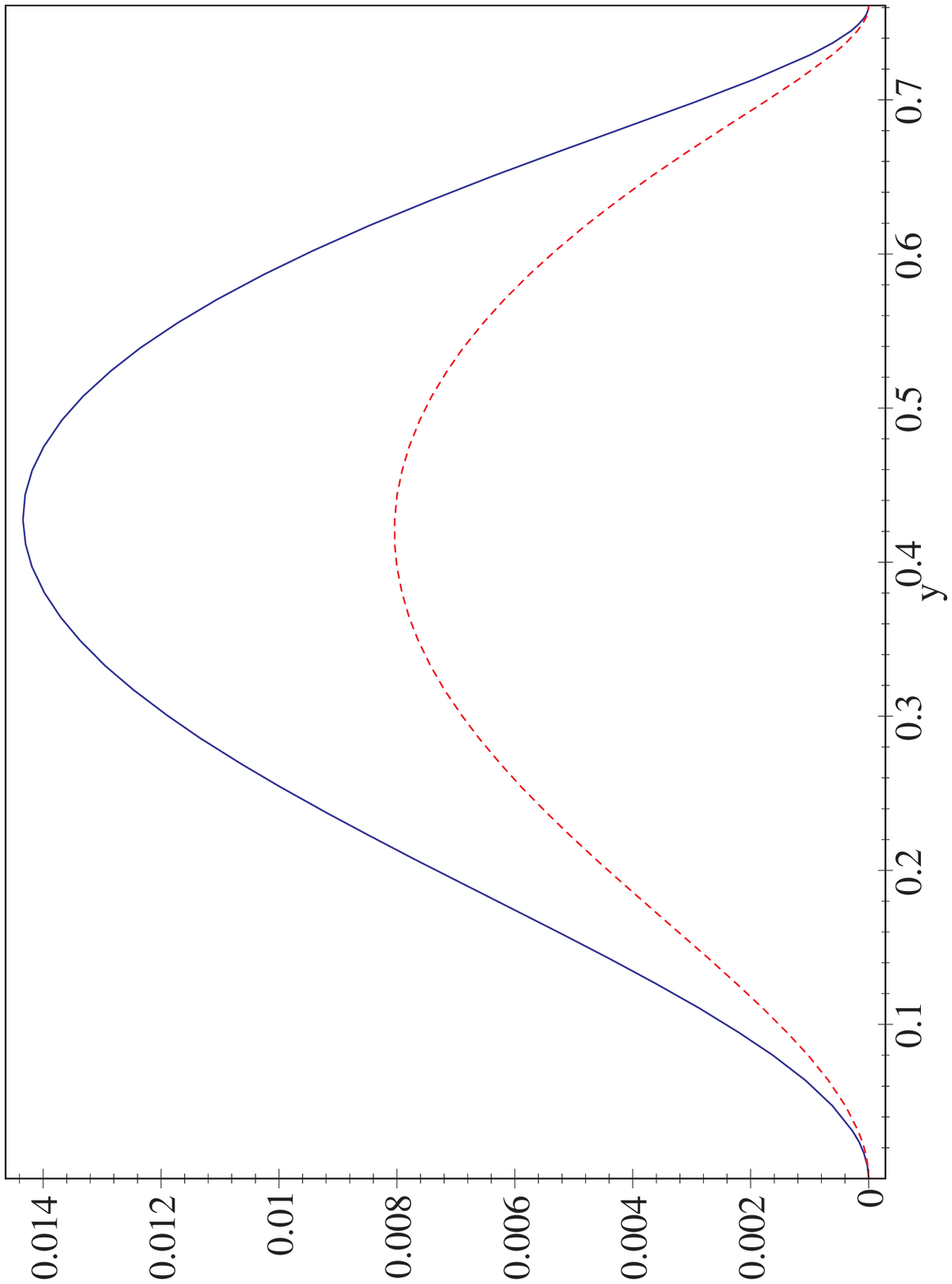}
\end{turn}
}
\caption{Electron spectra
$\left({1}/{\Gamma_0}\right)\left(
{{\rm d}\Gamma}/{{\rm d
} y}\right)$ for the $B\to D'e\nu$ decay
as a function of the rescaled lepton energy $y=2E_e/m_B$.
 Dashed curves show the $m_Q\to\infty$ 
limit, solid curves include first order $1/m_Q$ corrections.} 

\end{figure}
\begin{figure}
\centerline{
\begin{turn}{-90}
\epsfxsize=14 cm
\epsfbox{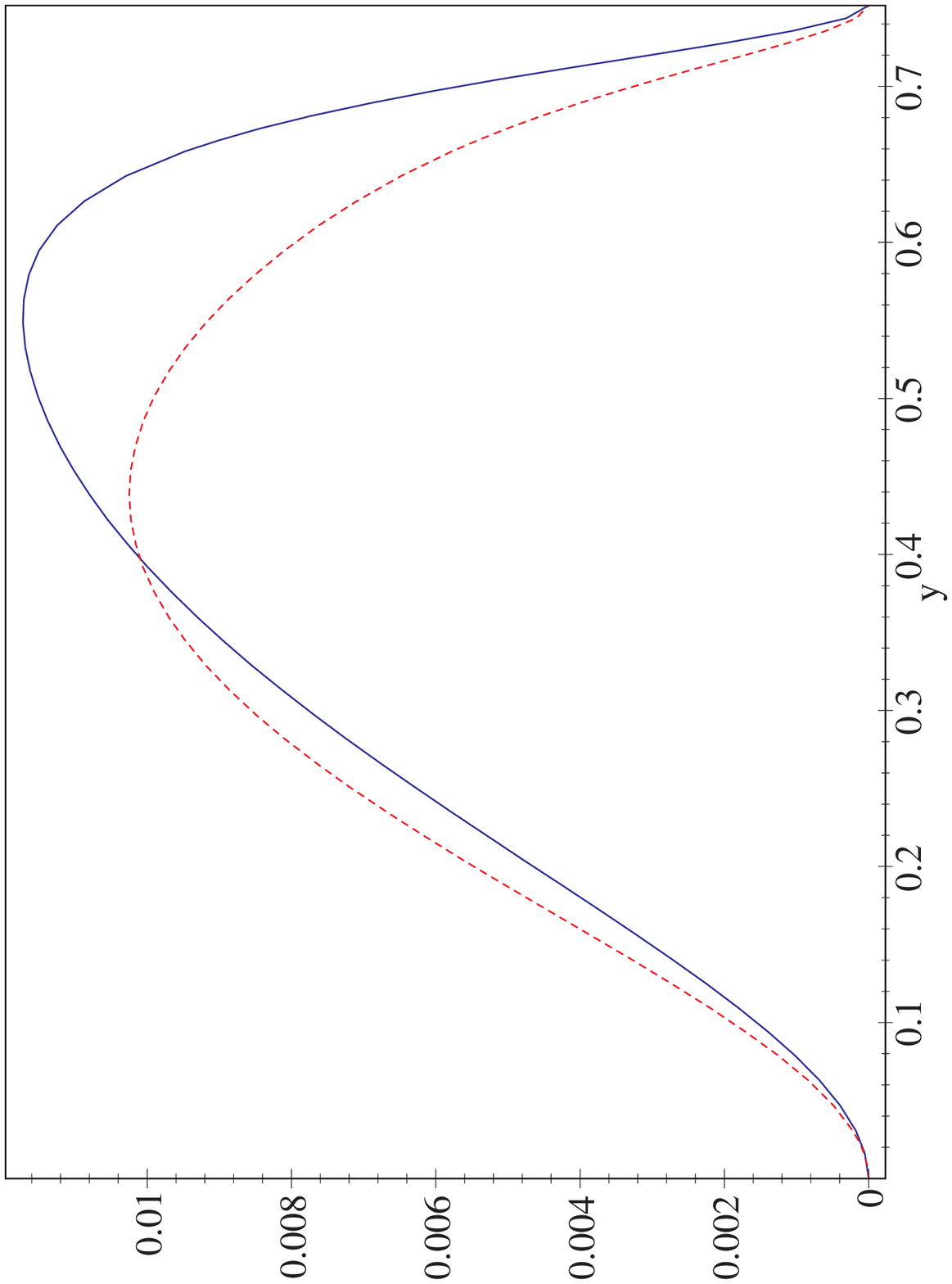}
\end{turn}
}
\caption{Electron spectra
$\left({1}/{\Gamma_0}\right)\left(
{{\rm d}\Gamma}/{{\rm d
} y}\right)$ for the $B\to D^{*}{'}e\nu$ decay
as a function of the rescaled lepton energy $y=2E_e/m_B$.
 Dashed curves show the $m_Q\to\infty$ 
limit, solid curves include first order $1/m_Q$ corrections.} 

\end{figure}


\begin{references}
\bibitem{cleo} CLEO Collaboration, A. Anastassov {\it et al.}, 
Phys. Rev. Lett. {\bf 80}, 4127 (1998).
\bibitem{aleph} ALEPH Collaboration,  D. Buskulic {\it et al.}, 
Z. Phys. C  {\bf 73}, 601 (1997).
\bibitem{opal} OPAL Collaboration,  R. Akers {\it et al.}, Z. Phys. C {\bf
67}, 57
(1995).
\bibitem{iw} N. Isgur and M.B. Wise, Phys. Lett. B {\bf 232}, 113 (1989);
{\bf 237}, 527 (1990); M.B. Voloshin and M.A. Shifman,
Sov. J. Nucl. Phys. {\bf 45}, 292 (1987); {\bf 47}, 511 (1988).
\bibitem{iw1} N. Isgur and M.B. Wise, Phys. Rev. D {\bf 43}, 819 (1991).
\bibitem{llsw} A.K. Leibovich {\it et al.}, Phys. Rev. D {\bf 57}, 308 (1998).
\bibitem{orb} D. Ebert, V.O. Galkin  and R.N. Faustov, Phys. Rev. D {\bf
61}, 014016 (2000).
\bibitem{egf} D. Ebert, V.O. Galkin  and R.N. Faustov, Phys. Rev. D {\bf
57}, 5663 (1998);
 {\bf 59}, 019902 (1999) (Erratum).
\bibitem{fgm} R.N. Faustov, V.O. Galkin and A.Yu. Mishurov, Phys.
Lett. B {\bf 356}, 516 (1995); {\bf 367}, 391 (1996)  (Erratum);
Phys. Rev. D {\bf 53}, 6302 (1996); R.N. Faustov and V.O. Galkin,
Phys. Rev. D {\bf 52}, 5131 (1995).
\bibitem{efg} D. Ebert, R.N. Faustov and V.O. 
Galkin, Phys. Rev. D {\bf 56}, 312 (1997).
\bibitem{fg} R.N. Faustov and V.O. Galkin, Z. Phys. C {\bf 66}, 119 (1995).
\bibitem{luke} M.E. Luke, Phys. Lett. B {\bf 252}, 447 (1990).
\bibitem{n} M. Neubert, Phys. Rept. {\bf 245}, 259 (1994);  Nucl. Phys. B
{\bf 416}, 786 (1994).
\bibitem{mr} T. Mannel and W. Roberts, Z. Phys. C {\bf 61}, 293 (1994).
\bibitem{falk} A.F. Falk, H. Georgi, B. Grinstein and M.B. Wise, Nucl. Phys.
B {\bf 343}, 1 (1990); A.F. Falk, Nucl. Phys. B {\bf 378}, 79 (1992).
\bibitem{3} A.A. Logunov and A.N. Tavkhelidze,  Nuovo Cimento  {\bf 29},
380  (1963).
\bibitem{4} A.P. Martynenko and R.N. Faustov, Teor.
Mat. Fiz. {\bf 64}, 179 (1985) [Theor. Math. Phys. {\bf 64}, 765 (1985)].
\bibitem{mass} V.O. Galkin, A.Yu. Mishurov and R.N. Faustov, Yad. Fiz.
{\bf 55}, 2175 (1992)  [Sov. J. Nucl. Phys. {\bf 55}, 1207 (1992)];
D. Ebert, R.N. Faustov and V.O. Galkin, hep-ph/9911283.
\bibitem{pot} D. Ebert, R.N. Faustov and V.O. Galkin,  Eur. Phys. J. C
{\bf 7}, 539 (1999).
\bibitem{gf} V.O. Galkin and R.N. Faustov, Yad. Fiz. {\bf 44}, 1575 (1986) 
 [Sov. J. Nucl. Phys. {\bf 44}, 1023 (1986)]; V.O. Galkin, 
A.Yu. Mishurov and R.N. Faustov, Yad.  Fiz. {\bf 51}, 1101 (1990) 
[Sov. J. Nucl. Phys. {\bf 51}, 705 (1990)].
\bibitem{schn} H.J. Schnitzer, Phys. Rev. D {\bf 18}, 3482 (1978).
\bibitem{buch} W.Buchm\"uller, Phys. Lett. B {\bf 112}, 479 (1982).
\bibitem{f} R.N. Faustov, Ann. Phys. {\bf 78}, 176 (1973); Nuovo
Cimento A {\bf 69}, 37 (1970).
\bibitem{delphi} DELPHI Collaboration, P. Abreau {\it et al.}, Phys Lett. B
{\bf 426}, 231 (1998); for review of experimental data on excited $B$ and
$D$ spectroscopy see V. Ciulli, hep-ex/9911044.

\end{references}
   \end{document}